\documentclass[aps,prd,amsmath
,eqsecnum            
,preprintnumbers
,nofootinbib         
,twocolumn         
]{revtex4}
\usepackage[dvips]{graphicx}
\usepackage{amsmath}
\usepackage{amsfonts}
\usepackage{amssymb}
\usepackage{longtable}   

\allowdisplaybreaks[4]     

\newcommand{\av}[1]{\langle{#1}\rangle}

\begin{document}

\title{Intermediate homogenization of the Universe and the problem of gravitational entropy}

\author{Krzysztof Bolejko}
\affiliation{Sydney Institute for Astronomy, 
The University of Sydney, New South Wales 2006, Australia}
\email{bolejko@physics.usyd.edu.au}

\author{William R. Stoeger}
\affiliation{Vatican Observatory Research Group, University of Arizona, Tucson, Arizona 85721, USA}

\begin{abstract}
This paper studies intermediate homogenization of inhomogeneous cosmological models. It shows
that spherically symmetric models, regardless of the equation of state, can undergo
intermediate homogenization, i.e. a model can approach a homogeneous and
isotropic state (which acts as a saddle point) from a relatively wide range of
initial inhomogeneous conditions. 
The homogenization is not permanent -- just temporary. Eventually the model evolves toward a future inhomogeneous state.
We also looked at the problem of the gravitational entropy. All definitions of entropy
that we checked give decreasing gravitational entropy during the homogenization process.
Thus, we should either accept that gravitational entropy can decrease or try to define 
it in other ways than just via density gradients, as these decrease during homogenization.
\end{abstract}

\maketitle

\section{Introduction}

Cosmological observations seem to suggest that on large-scales the Universe is homogeneous, at least in some statistical sense.
The strongest argument comes from the isotropy of the cosmic microwave background radiation (CMB). The Ehlers-Geren-Sachs (EGS) theorem \cite{EGS} and the `almost EGS theorem' \cite{SME} imply that if anisotropies in the CMB are small for all fundamental observers then locally the Universe is almost spatially homogeneous and isotropic. In addition, if there were large inhomogeneities in the Universe they would manifest themselves in CMB temperature fluctuations via the Rees--Sciama effect \cite{RS68,IS06,IS07}. Moreover, the  success of the homogeneous 
Friedmann--Lema\^itre--Robertson--Walker (FLRW) models in describing cosmological observations seems to provide further evidence for the large scale homogeneity of our Universe ---  cosmological observations are successfully analyzed within the framework of homogeneous models, including supernova Ia luminosity distance observations \cite{sn1,sn2}, baryon acoustic oscillations \cite{bao} and the CMB \cite{cmb}.

However, when extrapolating the present-day cosmological model back to early times it becomes apparent that 
the size of causally connected regions becomes smaller and smaller.
Thus, if the present-day observed Universe were not causally connected
in the past then it seems very possible that it was highly inhomogeneous in its early stages \cite{W84,E06}. 
Hence, we have the following  questions: {\em how is it possible that the Universe is homogeneous}? {\em was it homogeneous from the very beginning}?
or {\em did the homogeneity develop due to some kind of dynamical process}? and if so then {\em what kind of processes were responsible for this}?

Over past decades cosmologists invented several approaches to explain the large-scale homogeneity.
In 1960s Misner proposed the idea of {\em chaotic cosmology} \cite{Misn1968,Misn1969},
in which the present-day large-scale homogeneity and isotropy of the Universe developed in the course of time. The initial state could have been quite chaotic.
Misner considered the Bianchi type I model with viscosity. Among homogeneous and anisotropic models
the Bianchi I models are the simplest and in the case of vanishing shear they reduce to the parabolic FLRW solution.
However, Collins \& Hawking \cite{CH73} showed that among homogeneous and anisotropic models
those that approach a homogeneous and isotropic state as $t \to \infty$  are of measure zero. 
Bonnor and Tomimura \cite{B74,BT76} showed that 
inhomogeneous cosmological models with Lema\^itre-Tolman \cite{L33,T34} and Szekeres \cite{S75} geometries become homogeneous when one of the functions defining the model is exactly of a specific form.
As later confirmed by Silk \cite{S77} and Pleba\'nski \& Krasi\'nski \cite{PK06} such models are characterized by being free of curvature perturbations. Thus, they only possess decaying modes \cite{S77,PK06} (recently these decaying modes have been carefully treated by Wainwright and Andrews \cite{WaAn2009} -- see their Eq.(38)). Hence, these models also appear to be very special compared to a general set of inhomogeneous models. There is another category 
of expanding models, which approach a de Sitter configuration \cite{HaMo1982}.
These constitute an open set (for an explicit examples see 
\cite{WEUW2004}). In the case of the  Lema\^itre-Tolman models the Milne model is yet another possible future asymptote \cite{WaAn2009}.

This asymptotic behavior, however, does not seem to be able to account for the 
large-scale near homogeneity of the Universe, which seems to be a cosmic feature at least since the time of last scattering.
 Therefore, the conceptually different approach of {\em quiescent cosmology} was 
introduced by Barrow at the end of 1970s \cite{BaMa1977,Barr1978}.
Within this framework the Universe starts from a fairly homogeneous state.
An argument for quiescent cosmology is usually related to cosmic entropy
considerations. As the entropy always increases with time, the entropy of the early Universe must have been small. A simple estimation of the entropy of the early universe
seems to suggest that it must have been close to homogeneous -- otherwise its entropy would have been too large. Later on Penrose put forward the {\em Weyl curvature hypothesis} \cite{Penr1979}, which relates the gravitational entropy to the Weyl curvature
(we shall examine this connection later).
In this scenario, the initial state of the universe should be of vanishing Weyl curvature.
Later on Goode and Wainwright \cite{GoWa1985} showed 
that one can impose a weaker condition to explain homogeneity: the Weyl curvature should be small compared to Ricci curvature -- in the early Universe Ricci curvature should dominate over Weyl curvature (for a discussion see \cite{ErSc2002,HoSc2008}).

An alternative approach, called {\em inflation}, 
was proposed at the beginning of 1980s
(for a discussion see \cite{L90,VM05}). Starting from some
small homogeneous patch inflation leads to a very rapid exponential increase of its size through
the action the vacuum energy of a scalar field. Its very high negative pressure generates a
repulsive gravitational field.  A small 
patch would be inflated to a size much larger than the cosmic horizon. 
However, at the beginning of 1990s it was realized that inflation
cannot begin if the degree of inhomogeneity is too large.
If the metric of the spacetime is homogeneous and isotropic with all inhomogeneities dumped into the scalar field alone
then such model may undergo inflation \cite{SS87}. Also in the framework of linear perturbations around
the FLRW spacetime it was found that a suitable scalar field can initiate inflation \cite{AM11}.
Too large a degree of inhomogeneity prevents the onset of inflation.
Actually in order to start, inflation requires a homogeneous patch of at least the horizon size \cite{GP89,GP90,GP92,GS92}.
This means that we need some homogeneity to begin with and that,
without some other process producing that homogeneity, inflation is rather unlikely to occur in the real Universe.
\footnote{See Penrose\cite{Pen1989} for a very penetrating and readable account of the shortcoming of inflation. For a debate on pro and cons of inflation see \cite{Stein2009}.}

At the end of 1990s, using a dynamical-systems approach a very interesting feature was observed.
Studying Bianchi models it was found that most of the Bianchi models undergo  {\em intermediate isotropization} \cite{WE97a,WE97b}. 
Their phase spaces contain the Einstein-de Sitter model as a saddle point.
Hence, during the course of evolution, starting from a wide range of initial conditions, the system approaches the flat Friedmann model. Thus the model becomes almost isotropic.  Later, however, the system moves away from this state and becomes anisotropic again.

This paper aims to investigate whether inhomogeneous models possess a similar feature -- if {\em intermediate homogenization} can occur, and, if so,  then under what conditions.
Since the general case is extremely difficult, we shall assume spherical symmetry. If this mechanism occurs within spherically symmetric models then it is possible that it
may occur within a more general class of models. On the other hand
if intermediate homogenization does not happen within 
inhomogeneous spherical symmetric models, it is very unlikely that it happens within general asymmetric models.
If it does occur, intermediate homogenization would provide
a link between the chaotic and the quiescent approaches to cosmology.
Also if it turns out that intermediate homogenization is 
a generic phenomenon it would be a 
solution to the inflation problem. Inflation could occur once the model
approaches, in phase space, a homogeneous configuration. 

In Sec. II we write down the metric and the Einstein field equations for 
general spherically symmetric cosmological models along with a characterization
of cosmic matter. There follows in Sec. III a presentation of the results of
the numerical integration of six spherically symmetric inhomogeneous models
during a very early era of the universe demonstrating the rapid intermediate
homogenization they exhibit. This occurs as long as the spatial curvature $E$
is much less than the Ricci curvature. In Sec. IIID we
discuss various candidates for adequately representing gravitational entropy,
to determine if any of them might be positive-definite and monotonically increasing
with time, thus realizing Penrose's {\it Weyl-curvature hypothesis} \cite{Penr1979}.
We find that none of the candidates we examine do so. We give a summary of our conclusions in Sec. IV.

\section{Spherical symmetric models}

The most general form of a spherically symmetric metric is

\begin{equation} 
{\rm d} s^2 = -{\rm e}^{A}{\rm d} t^2
+ +\frac{R'^{2}}{1+2E} {\rm d}  r^2 + R^2 {\rm d} \vartheta^2 + R^2 \sin^2 \vartheta {\rm d} \varphi^2, \end{equation}
where the functions $A,R,$ and $E$ depend on $t$ and $r$, the prime denotes the partial derivative with respect to $r$, $R' \equiv \partial R / \partial r$.
We presume that the mass-energy which sources the gravitational field can be represented by a fluid. In several of
our models it will be an imperfect radiative fluid. In several others we shall specialize it to a perfect fluid with
a dust ($p = 0$) equation of state. From the Einstein equations we obtain the
following evolution equations \cite{BL08,LaBo2010}

\begin{eqnarray}
&& \dot{R}^{2}=2E + \frac{2M}{R}+ \frac{1}{3}\Lambda R^2, \label{Rdot} \\
&& \dot{M} = -\frac{1}{2} \kappa p \dot{R}R^{2},\label{Mdot} \\
&& \dot{E} = \frac{A'}{2}  (1+2E) \frac{\dot{R}}{R'}, \label{Edot}
\end{eqnarray}
where $\kappa = 8 \pi G /c^4$, $G$ is the gravitational constant, $c$
is the speed of light, and  $\Lambda$ is the cosmological constant --
for the early universe epoch considered in this paper the cosmological constant
is completely negligible, and so we set it to zero --
$p$ is pressure, and a dot denotes the partial derivative with respect to proper time, 
$\dot{R} \equiv {\rm e}^{-A/2} \partial R / \partial t$.
The gradient of the function $A$ follows from $T^{ab}{};_b =0$ and is
\begin{equation}  \frac{A'}{2} = \frac{-p' + \frac{2}{\sqrt{3}} ( \lambda \sigma)'
+2 \sqrt{3} \lambda \sigma R'/R 
}{\rho+p}, 
\label{Aprim}
\end{equation}
where $\lambda$ is the viscosity coefficient, $\sigma$ is the scalar of the shear 
$\sigma^2 = \sigma_{a b}\sigma^{a b}/2$, and we have employed Eckhart's model for treating
the viscous stress\cite{Eck1940}. $\rho$ is the energy density and is given by
\begin{equation}
\kappa \rho = \frac{2M'}{R^2R'} \label{rho}.
\end{equation}

\section{Results}

In \cite{BoSt2010} we argued that inhomogeneous models can undergo the process of homogenization starting from a set of initial conditions that do not have to be of measure zero.
In \cite{BoSt2010} we provided only qualitative 
or semiquantitative arguments, whereas here we provide
a more detailed analysis and explicit examples.

\subsection{Set-up}

The solution  algorithm used in our numerical calculations consists of following steps:
\begin{enumerate}
\item
The radial coordinate is chosen to be the areal radius at the initial instant:
$\bar r = R(t_{i},r)$. However, to simplify the notation we will omit the bar
and denote the new radial coordinate by $r$.
\item
The initial instant is set to be
when the energies in the universe models are around those of the GUT era, i.e.  $10^{16}$ GeV.
For radiative models this implies 
temperature around $2.5 \times 10^{31} K$ and 
an energy density of $4.5 \times 10^{110}$ J/m$^3$. \footnote{
If earlier times were properly described by a flat Friedmann
model dominated by radiation, then this moment would correspond to 
$ t = (6\pi G \rho)^{-1/2} \approx
4 \times 10^{-41}$ s after the Big Bang (where $\rho = \epsilon / c^2$,
and $\epsilon = 4.5 \times 10^{110}$ J/m$^3$). This is how the cosmic age is calculated in the standard cosmology. However, as for calculations presented here the exact value of the  $t_i$ is not important, we do not specify it -- what is important is the exact value of the initial energy density.}

\item
The initial energy density profile is assumed to be
\[ \rho_i = \rho_0 \left(1+ \frac{\beta}{1+(r/\alpha)^2} \right), \]
where $\beta = 1000$, $\rho_0 = 2.05 \times 10^{101}$ J/m$^3$, and $\alpha = 2\times10^{-28}$ m.
This profile describes a single inhomogeneity of amplitude $A$. Thus, this
is a convenient profile to study the process of the homogenization.

\item
The initial mass follows from (\ref{rho}) by integration.
\item We consider 6 different models: SS1-3, LT1-3
(SS stands for a general spherically symmetric model,
while LT stands for the Lema\^itre--Tolman model,
which is a spherically symmetric solution with dust).
The initial values of the functions $E$, which really
represents the spatial curvature (see Eq. (A4) of
Wainwright and Andrews \cite{WaAn2009}), are
\begin{itemize}
\item models SS1 and LT1: $E = 10^{-4} \times \frac{2M}{R}, $
\item models SS2 and LT2: $E = 10^{-3} \times \frac{2M}{R}, $
\item models SS3 and LT3: $E = 10^{-2} \times \frac{2M}{R}. $
\end{itemize}
The reason for choosing $E$ as above is that in \cite{BoSt2010}
it was shown that a necessary condition for homogenization
is that $E\ll M/R$. This does not however mean that the 
model is spatially flat. Here for example 
the spatial curvature is more than 100 orders of magnitude larger than 
the present-day spatial curvature in 
the FLRW model with $\Omega_m = 0.3$ and $\Omega_\Lambda = 0$.
It is also important to remember that in general $E = E(t, r)$.

\item 
Unfortunately the question of the equation of state under extreme conditions 
remains unanswered. 
Although observations of neutron stars rule out
some equations of state, they still allow for a wide
range of possibilities \cite{DPRRH2010}. The ground-based experiments
on the quark-gluon plasma suggest that it behaves 
like a ``perfect'' liquid  with zero or at least very low viscosity \cite{LTAA2010}.
If this is true even with a very strong gravitational field, the question will most likely remain unanswered for 
years. Thus, we consider 2 types of equations of state in this paper:
one with large viscosity, and the very opposite scenario -- the dust equation of state.
The equation of state thus takes the following form:

\[ p = K \rho - \eta \rho \Theta. \]
The first part of the above equation of state is the standard 
barotropic part, the second part is due to viscosity.
The coefficient $\eta$ is the bulk viscosity (we 
also consider the shear viscosity, which
relates the anisotropic stress-tensor to shear
via $\pi_{ab} = - \lambda \sigma_{ab}$; see also (\ref{Aprim})).
The coefficient $\Theta$ is the expansion scalar
\[ \Theta = u^{\alpha}{}_{;\alpha} =
 \frac{\dot{B}}{2} + 2 \frac{\dot{R}}{R}, \]
where $u^a$ is the the velocity field, and ${\rm e}^B = R'^2/(1+2E)$.

The specific forms of the equation of state for models SS and LT are as follows: 
\begin{itemize}
\item models SS1, SS2, SS3: $K = \frac{1}{3}$, $\eta = 10^{-31}$ s,
(so we choose $K$ as that for an ultra-relativistic fluid) 
and $\lambda = 4 \times 10^{22}$ Pa-s [Pascal-seconds],
which is an extremely large value 
(for comparison: viscosity of water at 20 degrees C is
$\lambda = .001$ Pa-s, viscosity of motor oil is $0.25$ Pa-s,
and viscosity of pitch is $2.3 \times 10^8$ Pa-s).

\item models LT1, LT2, LT3: $K = 0$, $\eta = 0$, $\lambda=0$.
\end{itemize}
\item Given the initial conditions and the equation of state as above,
we solve the evolution equations (\ref{Rdot})--(\ref{Edot})
using the 4th order Runge--Kutta methods. The code was written in Fortran.
The results of the evolution and the discussion are presented in the next section.

\end{enumerate}

\subsection{Intermediate homogenization}

\begin{figure*}
\includegraphics[scale=0.43]{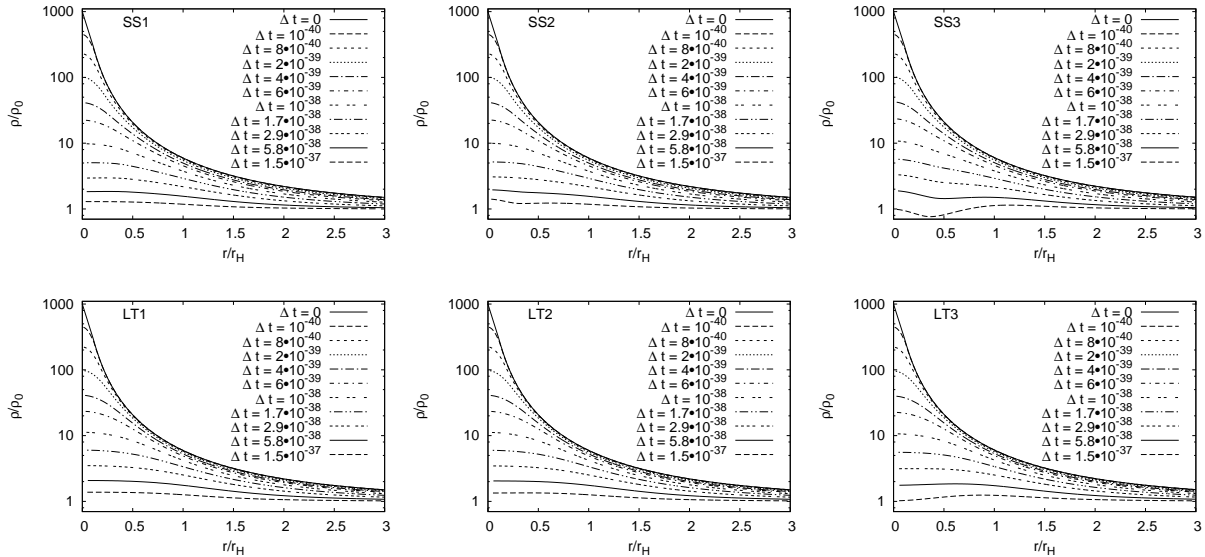}
\caption{Evolution of the models we have considered:
 density profile at the initial 
instant $10^{-38}$ s and afterwards.
$\rho_b$ is the background density at a given time instant, and $r_H$
is the Hubble radius at that time.}
\label{fig1}
 \label{inhom}
\end{figure*}

\begin{figure}
\hspace{0.6cm}
\includegraphics[scale=0.4]{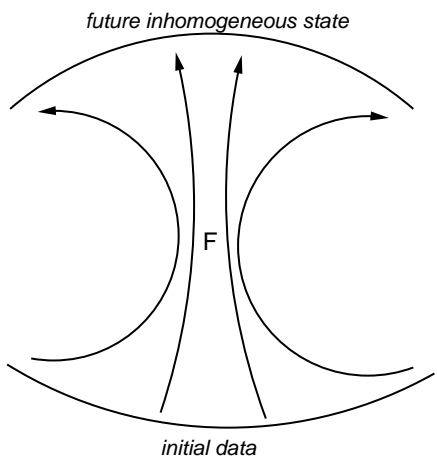}
\caption{Schematic presentation of the evolution of the model we are considering.
F is the Einstein-de Sitter model that acts like a saddle point.}
 \label{fig2}
\end{figure}

Figure \ref{fig1} shows our results.
The radial dependence is expressed in terms of the Hubble radius\footnote{The Hubble radius 
is expected to be of a similar magnitude as the distance to the horizon, for example,
for the present day standard cosmological model, $r_H \approx 4.2$ Gpc,
while the present-day distance to the horizon is roughly $14.2$ Gpc.}, which is defined as 
\[ r_H = \frac{c}{H}, \]
where $H$ is the Hubble parameter. 
Figure. \ref{fig1} exhibits the results of the 6 models we studied.
At the initial instant the background energy 
density is $2.05 \times 10^{101}$ J/m$^3$ (or in natural units $10^{16}$ Gev),
which in a radiation-dominated homogeneous model would translate to 
$t_i \approx 1.85 \times 10^{-38}$ s. However, as we 
have already mentioned, to 
define the model and calculate its evolution
we do not need to know the exact value of $t_i$. What is needed is just the
amount of time elapsed after $t_i$. This is expressed by $\Delta t$.
Thus, Figure \ref{fig1} presents snapshots of the evolution at different $\Delta t$,
whose values are shown in the top right corner of each panel.
The vertical axis shows the ratio of the local density to the density of a background homogeneous model.
As seen in all the models, the initial inhomogeneity damps away relatively quickly, leading to intermediate homogenization. 
The time-scale is of order of the Hubble time-scale (i.e. $\sim$ age of the
Universe). As the Universe is young, the whole process
proceeds quickly.
However, in the SS3 and LT3
models, where the initial spatial curvature was larger than in the SS2/LT2 and SS1/LT1 models,   
a new inhomogeneity begins to appear at $\Delta t \approx 10^{-37}$s.
This is because, roughly speaking, the spatial curvature
evolves as $E R^{-2}$ while the energy density as $M R^{-3}$.
Thus it takes a bit longer in models SS2/LT2 and SS1/LT1 before spatial
curvature becomes dominant and the model starts to evolve towards a future inhomogeneous state.
This is schematically presented in Fig. \ref{fig2}.
A model can start from a wide range of initial conditions, and as long
as $E \ll M/R$ it  undergoes intermediate homogenization. Later
on, however, the spatial curvature term starts to dominate and the model
becomes inhomogeneous again.
This is very similar to the intermediate isotropization observed in the Bianchi models
\cite{WE97a,WE97b}, where the flat Friedmann model (F on Fig. \ref{fig2}) 
acts as a saddle point. Also a similar feature was observed in the case
of the silent models, in particular the Szekeres model \cite{WE97c}.

Here we have studied not only dust models but also viscous-fluid models, and we found little difference between these two cases. From
what we have just discussed above, intermediate homogenization is really
an effect linked to the dynamics of the geometry.

\begin{figure*}
\hspace{0.6cm}
\includegraphics[scale=0.67]{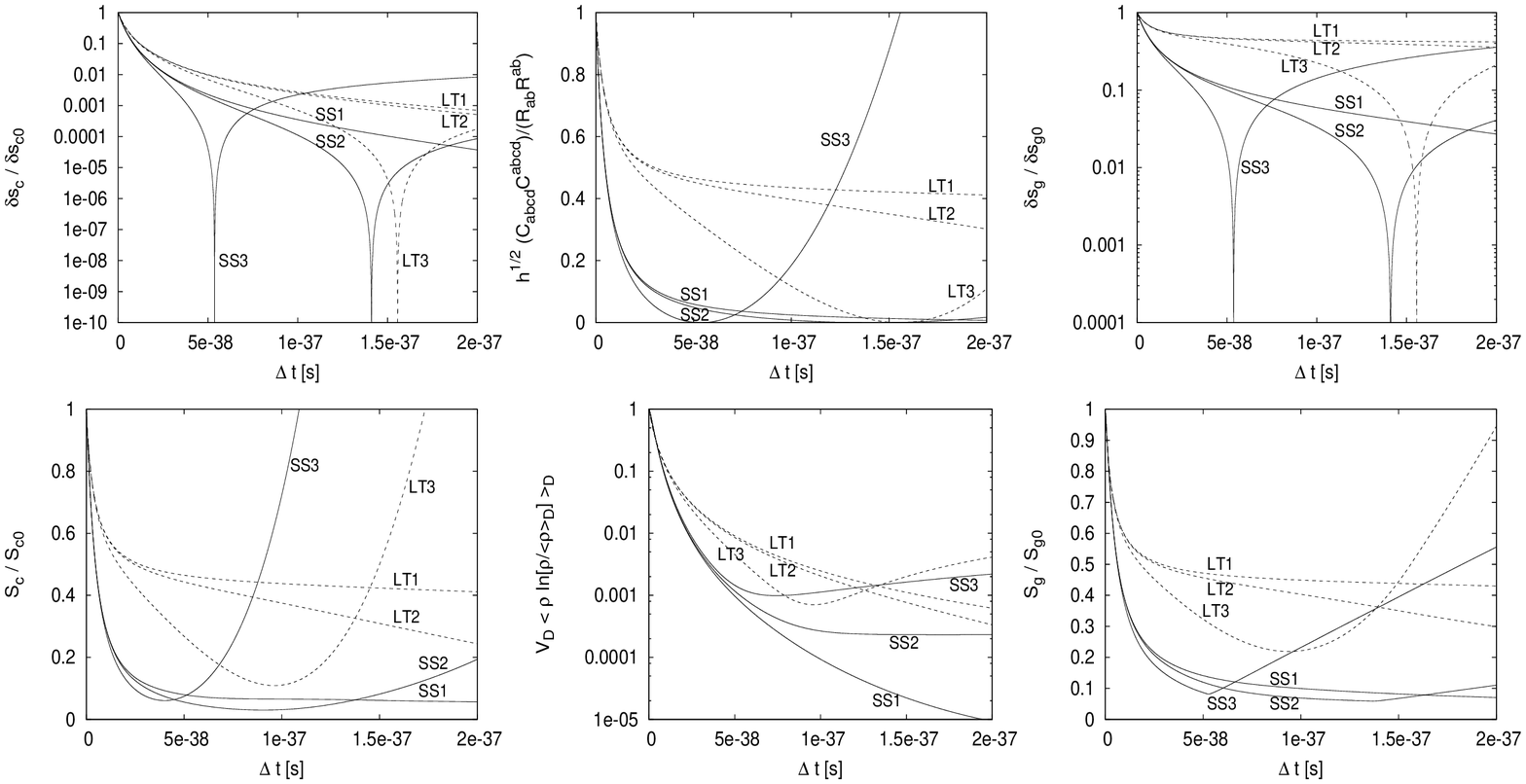}
\caption{Different representations of the gravitational entropy and their evolution.
The entropy was calculated at $r_i = 0.75 r_H$. For the integrated version
the domain is $r<r_i = 0.75 r_H$. The gravitational entropy is scaled so that
it is 1 at the initial instant.}
 \label{fig3}
\end{figure*}

\subsection{Domination of decaying modes}

A convenient way of thinking about intermediate
homogenization  is to describe it in terms of decaying modes. This relies on being able to decompose the evolution of density into decaying and growing modes.
Since the decaying modes are always 
decreasing in amplitude, and the growing modes increasing, at later times the growing modes will
dominate, whereas at very early times, the decaying modes will be much, much larger than the growing modes.
If the past is dominated by a decaying mode, then naturally the universe must undergo  a period of homogenization, where inhomogeneities decay, and the growing modes are still very small. Depending on the relative rates of the decay and growth of these modes, we may have a longer or shorter period of homogenization.
Although the above reasoning is helpful in understanding the process of homogenisation,
we should bear in mind that in the nonlinear regime, 
one cannot simply decompose the evolution
 onto a linear combination of  growing and decaying modes \cite{suss2013a}.
Also some configurations do not have solutions 
in terms of growing and decaying modes, but only in terms of oscillatory modes.

\subsection{The Gravitational Entropy}

A single-component fluid should obey the Gibbs-Duhem relation

\begin{equation}
{\rm d} U + p {\rm d} V = T {\rm d} S,
\label{gibbs}
\end{equation}
where $U$ is the internal energy, $p$ is the pressure, $V$ is the volume, $T$ is temperature, and $S$ is the thermodynamic entropy.
Introducing the particle number density $n$ we can write
equation (\ref{gibbs}) as
 
\begin{equation}
\label{gibbs2}
{\rm d} (\epsilon/n) + p {\rm d} (1/n)  = T  {\rm d} S : = \omega.
\end{equation}
If $\omega$ has an integration factor 
\[ \omega \wedge {\rm d} \omega = 0, \]
then (\ref{gibbs2}) can be solved for $T$ and $S$.
The above is equivalent to

\[ {\rm d} \epsilon  \wedge {\rm d} p  \wedge {\rm d} n  = 0. \]

This is always fulfilled if at least one of the following conditions holds: (1) the fluid is static,
 (2) the spacetime possesses a
high degree of symmetry -- isometry of symmetry groups must have orbits of dimension 
at least 2 (for example spherical symmetry, so physical quantities depend on at most
two coordinates), (3) the fluid's equation of state is barotropic,  $p= p (\epsilon)$.
In other cases a solution may not exist. For a detailed discussion and examples see \cite{QuHe1995,KrQS1997,PK06}.

In this paper we assume a barotropic equation of state, so that this thermodynamic scheme exists (i.e. $\omega$ has an integration factor). In this case  eq. (\ref{gibbs}) reduces to  \cite{E71}

\begin{equation}
\rho T \dot{S} = \lambda \sigma_{ab} \sigma^{ab}.
\end{equation}
As we see, the change in the thermodynamic entropy is always positive.

However, as noted by Penrose, in the absence of gravitation, a homogeneous state is a state of maximal entropy, whereas in the presence of gravitation we observe that the
natural tendency is  for the system to evolve from a state of 
homogeneity to states of greater clumpiness.
Thus, there have been attempts to define the gravitational entropy.
As homogeneous and isotropic models are of zero Weyl curvature
a natural choice is to relate the gravitational entropy with  Weyl curvature.
Here we study the following quantities:
\begin{enumerate}
\item The standard canonical definition, i.e. the ratio of the Weyl to Ricci curvature \cite{WaAnd1984}

\begin{equation}
\delta s_c = \frac{C_{abcd} C^{abcd}}{{\cal R}_{ab} {\cal R}^{ab}},
\end{equation}
where $C_{abcd}$ is the Weyl tensor, and  ${\cal R}_{ab}$ is the Ricci tensor.
\item The integrated version of the canonical definition above. 

\begin{equation}
S_c = \int {\rm d}^3x \sqrt{h} \delta s_c
\label{ISc}
\end{equation}

\item Following \cite{GrHe2001}, we take the canonical definition
multiplied by the square root of the determinant of the spatial metric

\begin{equation}
S_h = \sqrt{h} \frac{C_{abcd} C^{ab}}{{\cal R}_{ab} {\cal R}^{ab}} .
\label{S_GrHe}
 \end{equation}

\item Following \cite{HoBM2004,Suss2013} we examine the following quantity
\begin{equation}
S_a = V_{\cal D} \av{ \rho \ln \frac{\rho}{\av{\rho}_{\cal D}} }_{\cal D}, 
\label{ISb}
\end{equation}
where $\av{}_{\cal D}$ is the volume average over the domain ${\cal D}$.

\item 
Following \cite{CET13} we examine the following formula
for the gravitational entropy
\begin{equation}
S_g = \int_V \delta s_g = \int {\rm d}x^b {\rm d}x^c {\rm d}x^d \eta_{abcd} z^a \frac{\rho_{grav}}{T_{grav}}
\label{ISg}
\end{equation}
where $z^a$ is a spacelike unit vector aligned with the Weyl principal tetrad, 
$\rho_{grav}$ and $T_{grav}$ are effective energy density and temperature of the gravitational field respectively

\[\rho_{grav}  \sim \sqrt{T_{abcd} u^a u^b u^c u^d}, \]
and 
\[ T_{grav} \sim | \frac{1}{3} \Theta + \sigma_{ab} z^a z^b  + \dot{u}_a z^a |, \] 
where $T_{abcd}$ is the Bel--Robinson tensor.

\end{enumerate}

For each of our models we calculate the gravitational entropy
as defined above. We arbitrarily choose a shell
of radial coordinate $\tilde{r}$, which at the initial instant  
is equal to $\tilde{r}= 0.75 r_H$ -- as seen from Fig. \ref{fig1}.
This corresponds to a transition region between the central peak and the almost homogeneous tail.
When integration is involved, i.e. definitions (\ref{ISc}),
(\ref{ISb}), and (\ref{ISg}) we consider a domain of $r\le \tilde{r}$. 

The results are presented in Fig. \ref{fig3}.
Regardless of the definition, there is always a period of time where gravitational entropy
decreases. 
In terms of (\ref{S_GrHe}) it was already noticed in  \cite{HoBM2004} that 
this quantity does not need to be monotonically increasing.
This should not be a surprise, as the above definitions incorporate, in one way or another,
the gradient of energy density. During homogenization the gradient decreases, and
so the gravitational entropy, which is based on it, also decreases.
This shows that other definitions, for example the family
of density contrast indicators presented and discussed in \cite{MeTa1999} (which in normal 
late-time cosmology increase, cf. \cite{Bole2007}) should also be decreasing
during intermediate homogenization,
which is in agreement with recent studies of gravitational entropy 
within the Lema\^itre-Tolman models \cite{RSJL2013}.

Unlike in thermodynamics, we do not have any fundamental theorem
to define the gravitational entropy. The above definitions are  based 
on some expectations rather than on solid theorems.
Thus we should either accept that gravitational entropy can decrease
or we should seek definitions that are not based on density gradients, or at least are defined in such a way that even during homogenization they increase.

\section{Conclusions}

It has been known for some time that in Bianchi models 
intermediate isotropization can occur. 
This paper has aimed to study if an analogous feature -- intermediate homogenization --
can occur in inhomogeneous cosmological models.
Our study has been confined to spherically symmetric models. We have shown that,
under specific conditions, an inhomogeneous system undergoes
homogenization. The homogenization can be permanent (special case)
or just intermediate (after some time the system becomes inhomogeneous again).
We have provided several examples -- some with bulk and shear viscosity,
and others with a pure dust equation of state. 

Obviously much more remains to be done. We have simply provided a number of different numerically
integrated examples showing that intermediate homogenization occurs at early times, whenever the
the spatial curvature is much less than the Ricci curvature. In the very early universe this
can be realized for a relatively broad class of initial conditions. 
However, we have not
directly addressed how generic this behavior is, nor how long these universes remain spatially homogeneous, nor
how close they come to an FLRW (spatially homogeneous) universe. This is a project for the future.

Intermediate homogenization may remind us of isotropic initial singularities \cite{GoWa1985,WEUW2004}.
Some of the these intermediate homogenization models may have isotropic initial singularities, and some
may not. Since the initial conditions cannot be set at the initial singularity itself, and since an
isotropic singularity in not generic for these models, there is no direct relationship between the
initial conditions leading to intermediate homogenization and isotropic initial singularities. It is
very likely that the very early universe emerging from the Planck era before the onset of inflation 
was spatially inhomogeneous. What our results indicate is that those primordial inhomogeneities are
at least partially dissipated during the period when the spatial curvature is 
very much less than the Ricci curvature generated by the mass-energy density. 
There would be a considerable range of initial
conditions for which this would be true.
 
We have also considered the problem of entropy generation. The thermodynamic entropy always increases,
or, in the case of a perfect fluid, is constant. We tried to evaluate the gravitational entropy.
We used several formulas that were based on the Weyl tensor, Bel-Robinson tensor, or the
averaging of the density filed. All of them were decreasing during the process of intermediate homogenization. 
Therefore, unlike the thermodynamical counterpart, the gravitational entropy
may not necessarily be a monotonically increasing quantity.
Indeed, if the gravitational entropy describes the degree of inhomogeneity
then it should decrease during the process of homogenization.  
Consequently, if the early universe was dominated by the decaying mode
then its gravitational entropy must have been decreasing with time.
The same holds for the period of inflation, during which the gravitational entropy should
also be decreasing with time.

The question remains whether these results are 
special to spherical symmetry or whether  
they apply as well to more general geometries.
For the dust case, i.e. the Szekeres model, or a very special generalization to viscous fluid 
(so that there is no acceleration), the Roy-Singh model \cite{RS82}, one can show that intermediate homogenization occurs  \cite{BoSt2010,WE97c}.
Still the Szekeres and Roy-Singh models are not fully general \cite{BCK2011}. Thus it would be interesting, and important, to know if it can occur in the real Universe.

\medskip

\medskip

\section*{Acknowledgments}

We thank Henk van Elst for his careful reading and critique of the manuscript, and for
a number of corrections and helpful suggestions.

\end{document}